\documentclass[RNAAS]{aastex631}
\usepackage{ulem}

\begin{document}

\title{Comparative study of low-temperature opacities with GARSTEC models}

\author[0000-0002-1383-4173]{Pedro Díaz Reeve}
\affiliation{Institute of Space Sciences (ICE, CSIC), Carrer de Can Magrans S/N, E-08193, Cerdanyola del Vallès, Spain}
\affiliation{Institut d’Estudis Espacials de Catalunya (IEEC), Carrer Gran Capità 2, E-08034, Barcelona, Spain}

\author[0000-0001-6359-2769]{Aldo Serenelli}
\affiliation{Institute of Space Sciences (ICE, CSIC), Carrer de Can Magrans S/N, E-08193, Cerdanyola del Vallès, Spain}
\affiliation{Institut d’Estudis Espacials de Catalunya (IEEC), Carrer Gran Capità 2, E-08034, Barcelona, Spain}

%\collaboration{20}{(AAS Journals Data Editors)}

%\author{F.X Timmes}
%\affiliation{Arizona State University}
%\affiliation{AAS Journals Associate Editor-in-Chief}

%\author{Amy Hendrickson}
%\altaffiliation{AASTeX v6+ programmer}
%\affiliation{TeXnology Inc.}

%\author{Julie Steffen}
%\affiliation{AAS Director of Publishing}
%\affiliation{American Astronomical Society \\
%1667 K Street NW, Suite 800 \\
%Washington, DC 20006, USA}

%% Note that the \and command from previous versions of AASTeX is now
%% depreciated in this version as it is no longer necessary. AASTeX 
%% automatically takes care of all commas and "and"s between authors names.

%% AASTeX 6.31 has the new \collaboration and \nocollaboration commands to
%% provide the collaboration status of a group of authors. These commands 
%% can be used either before or after the list of corresponding authors. The
%% argument for \collaboration is the collaboration identifier. Authors are
%% encouraged to surround collaboration identifiers with ()s. The 
%% \nocollaboration command takes no argument and exists to indicate that
%% the nearby authors are not part of surrounding collaborations.

%% Mark off the abstract in the ``abstract'' environment. 
\begin{abstract}

We present a comparative study of the effect of low-temperature opacities on stellar models up to the Red Giant branch (RGB), computed with the GARching STellar Evolution Code. We have used two sets of low-temperature opacities; \AE SOPUS (\AE) from the University of Padova and those from the Wichita State University group (F05). In the relevant range of temperatures for this study, $\log \kappa^{\AE}$ $<$ $\log \kappa^{F05}$. Therefore, to compare stellar evolutionary tracks, we performed a solar calibration of the $\alpha_{mlt}$, for each set of low-temperature opacities. After carrying such a calibration, we find that stellar evolutionary tracks are almost unaffected by the choice of low-temperature opacities, with largest variations of 25-30 K at the latest evolutionary stages of the RGB phase.

%however after carrying a mixing length $\alpha_{mlt}$ calibration with a solar model, the effective temperature of the resulting evolutionary tracks is not affected by the choice of low-temperature opacities.

\end{abstract}

%% Keywords should appear after the \end{abstract} command. 
%% The AAS Journals now uses Unified Astronomy Thesaurus concepts:
%% https://astrothesaurus.org
%% You will be asked to selected these concepts during the submission process
%% but this old "keyword" functionality is maintained in case authors want
%% to include these concepts in their preprints.
%\keywords{Classical Novae (251) --- Ultraviolet astronomy(1736) --- History of astronomy(1868) --- Interdisciplinary astronomy(804)}

%% From the front matter, we move on to the body of the paper.
%% Sections are demarcated by \section and \subsection, respectively.
%% Observe the use of the LaTeX \label
%% command after the \subsection to give a symbolic KEY to the
%% subsection for cross-referencing in a \ref command.
%% You can use LaTeX's \ref and \label commands to keep track of
%% cross-references to sections, equations, tables, and figures.
%% That way, if you change the order of any elements, LaTeX will
%% automatically renumber them.
%%
%% We recommend that authors also use the natbib \citep
%% and \citet commands to identify citations.  The citations are
%% tied to the reference list via symbolic KEYs. The KEY corresponds
%% to the KEY in the \bibitem in the reference list below. 

\section{1. Introduction} \label{sec:intro}

Several sets of Rosseland mean opacities for stellar interiors, with different physical inputs and conditions have been developed, such as The Opacity Project OP [\cite{Badnell_2005}] and OPAL [\cite{1996ApJ...464..943I}] among many others. In this research note we compared two sets of low-temperature opacities in the range 3.50 $\leq \log T \leq$ 4.50, which include molecules and dust as opacity sources in addition to atoms, and the impact they have in stellar evolutionary tracks for different masses, from the MS and up to the RGB phase. The scope of the comparison, encompasses the \AE SOPUS\footnote{Available at http://stev.oapd.inaf.it/cgi-bin/aesopus} web interface (Accurate Equation of State and OPacity Unit Software) [\cite{Marigo_2009}] and [\cite{Marigo_2022}] and the database set up by the Wichita State University group [\cite{2005ApJ...623..585F}], hereafter F05. A recent study of \AE SOPUS opacities on the AGB phase has been presented in [\cite{2022RNAAS...6...77C}]. Results presented here use the MB22 [\cite{Magg_2022}] solar mixture, but we tested that other chemical compositions such as GS98 [\cite{1998SSRv...85..161G}] and AGSS09 [\cite{2009ARA&A..47..481A}]
show similar results [\cite{diaz_reeve_pedro_2023_8108192}]. Stellar models were computed using GARSTEC [\cite{2008Ap&SS.316...99W}] version 20.1.

\section{2. Comparative Study}
\label{sec:methods}

\subsection{2.1. Preliminary Overview of the Opacity Data Sets}

%\cite{Marigo_2009} showed that differences in Rosseland mean opacity between \AE SOPUS and F05, $\Delta \log \kappa_{RM}$, in the range 3.50 $\leq \log T \leq$ 4.50 are comprised within $\pm$ 0.05 dex throughout the range -8.00 $\leq \log R \leq$ 1.00, except for narrow regions where differences reached $\pm$ 0.10/0.20 dex, for GS98 solar composition. Being $\log R$ defined as:
 
\begin{figure}[ht!]
%\plotone{rnaas.jpg}
\centering
\includegraphics[scale=0.46]{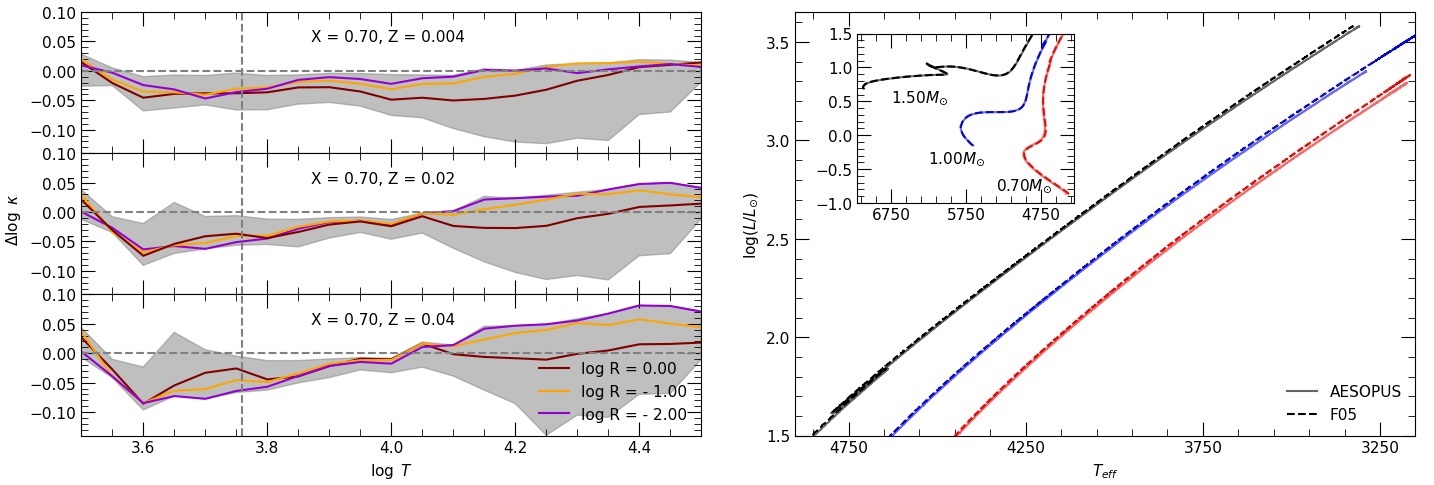}
\caption{Left: Comparison of the Rosseland mean opacity for \AE SOPUS and F05 for X = 0.70, Z = 0.004, 0.02 and 0.04. Gray vertical line shows the solar effective temperature $\log T \simeq$ 3.76, and gray shaded profile shows the range of variation for all $\log R$ values. Right: Evolutionary tracks for stellar models computed with GARSTEC with Z = 0.01998, and for $M/M_{\odot}$ = 0.70 (red track), 1.00 (blue track) and 1.50 (black track).}
\label{fig:general}
\end{figure}

Left panel in Figure \ref{fig:general} shows the differences in the Rosseland mean opacity between \AE SOPUS and F05, $\Delta \log \kappa = \log \kappa^{\AE} - \log \kappa^{F05}$, for MB22 chemical composition and for the cases X = 0.70 and Z = 0.004, 0.02 and 0.04, for temperatures between 3.50 $\leq \log T \leq$ 4.50, throughout the range -8.00 $\leq \log R \leq$ 1.00, where $\log R = \log \rho - 3 \log T + 18$. Such differences, in the range 3.50 $\leq \log T \leq$ 4.00 lie between +0.05/$-$0.08 dex, except for a peak at $\log T$ = 3.60 where differences reach up to $-$0.10 dex. For 4.00 $\leq \log T \leq$ 4.50 differences seem to be in a wider range between $\pm$ 0.10 dex and reaching at most $-$0.14 dex at $\log T$ = 4.25 for the case Z = 0.04. Differences appear to be larger with the increasing metallicity. Gray shaded regions show the differences for the whole $\log R$ range, while maroon, yellow and purple lines show differences for the cases $\log R$ = 0.00, -1.00 and -2.00 respectively, which span the $\log R$ values that better reproduce the approximate conditions in 0.70 - 1.50 $M_{\odot}$ stellar envelopes, up to the RGB phase.

\subsection{2.2. Solar Calibration} \label{subsec:solarcal}

We carried out solar calibrations using the \AE SOPUS  and F05 opacities. The chemical composition is almost independent of the choice of low-temperature opacities. Initial abundances are  $X_{\odot}$ = 0.70988 (0.70982), $Y_{\odot}$ = 0.27190 (0.27196) and $Z_{\odot}$ = 0.01822 (0.01822) for the \AE SOPUS (respectively F05) solar model. A more relevant difference appeared on the value of the $\alpha_{mlt}$ parameter, as its calibration in a solar model is sensitive to the choice of low-temperature opacities. We found $\alpha_{mlt}^{\AE}$ = 2.0530 for \AE SOPUS while for F05 the value was $\alpha_{mlt}^{F05}$ = 2.1487, i.e. $\alpha_{mlt}^{\AE} < \alpha_{mlt}^{F05}$. 

Near the solar surface at $\log T \simeq$ 3.76, \AE SOPUS shows smaller opacities than F05 (see left panel in Figure \ref{fig:general}), making the star more luminous (less opaque) and as a consequence, increasing its effective temperature. To compensate this change in temperature the mixing length parameter $\alpha_{mlt}$ decreases for less opaque models decreasing the effective temperature of the model star so that the solar effective temperature and luminosity are matched.

%% The "ht!" tells LaTeX to put the figure "here" first, at the "top" next
%% and to override the normal way of calculating a float position

\subsection{2.3. Comparative Study with GARSTEC models}

We computed stellar evolution models  for masses $M/M_{\odot}$ = 0.70, 1.00 and 1.50 and  $Z = 0.01998$, close to the solar calibrated $Z_{\odot}$. The $\alpha_{mlt}$ was used consistently with the low-temperature opacities as described above. Models extend up to the RGB phase. Right panel in Figure~\ref{fig:general} shows the computed evolutionary tracks.

Both sets of models were computed using low-temperature opacities for 3.20 $\leq \log T \leq$ 4.00, so the effect of molecules and dust, besides the atomic effects, are included in stellar evolutionary models, and OP opacities for $\log T \geq$ 4.10. 

Differences between models along the MS and the SGB phases are almost negligible for stellar evolutionary tracks for all masses, with differences of 10-15 K in the effective temperature of the model stars. Differences increase slightly at the latest stages of the RGB, reaching maximum values of  25-30~K at the RGB-tip. Such differences are originated mainly due to the fact that near $\log T \simeq$ 3.50, $\log \kappa^{\AE} > \log \kappa^{F05}$, i.e. there is a sign reversal in the opacity difference, so the effect of the solar calibration on $\alpha_{mlt}$ is no longer able to compensate for the opacity differences and effective temperature differences appear.

\section{3. DISCUSSION}

From the preliminary overview of the opacity data sets we found that differences between \AE SOPUS and F05 increase with metallicity and that the range of variation was wider for temperatures above $\log T$ = 4.00. However, mean differences between data sets were in the range $\pm$ 0.10 dex for temperatures 3.50 $\leq \log \leq$ 4.50 and 
for - 8.00 $\leq \log R \leq$ 1.00. Near the solar surface \AE SOPUS shows lower opacities than F05, which in a solar calibration was compensated by reducing the efficiency of the energy transport in near-surface convection, decreasing the value of the $\alpha_{mlt}$ in less opaque models. Results are presented for MB22 chemical composition, however GS98 and AGSS09 solar compositions show similar results. 

Stellar models for masses $M/M_{\odot}$ = 0.70, 1.00 and 1.50 and Z = 0.01998, showed differences around 10-15 K along the MS and SGB phases, while differences around 25-30 K appeared on the latest evolutionary phases of the RGB. We conclude that stellar models computed with low-temperature opacities either from \AE SOPUS or F05, are in agreement with each other when a solar calibration is carried out for $\alpha_{mlt}$. Results presented here are valid for stars on the MS, SGB and RGB evolutionary phases and should not be extrapolated to other evolutionary phases e.g. the AGB.

%% For this sample we use BibTeX plus aasjournals.bst to generate the
%% the bibliography. The sample631.bib file was populated from ADS. To
%% get the citations to show in the compiled file do the following:
%%
%% pdflatex sample631.tex
%% bibtext sample631
%% pdflatex sample631.tex
%% pdflatex sample631.tex

\bibliography{sample631}{}
\bibliographystyle{aasjournal}

%% This command is needed to show the entire author+affiliation list when
%% the collaboration and author truncation commands are used.  It has to
%% go at the end of the manuscript.
%\allauthors

%% Include this line if you are using the \added, \replaced, \deleted
%% commands to see a summary list of all changes at the end of the article.
%\listofchanges

\end{document}